
%
%
\input harvmac
\def\ie{{\it i.e.}}
\def\eg{{\it e.g.}}
\def\no{\noindent}
\def\o{\over}
\def\nl{\hfill\break}
\def\del{\delta}
\def\eps{\epsilon}\def\veps{\varepsilon}
\def\th{\theta}
\def\thb{{\th\kern-0.465em \th}}
\def\ZZ{{\bf Z}}

\def\csm{{\cal SM}}
\def\hc{\hat{\chi}} \def\hp{\hat{\phi}} \def\hD{\hat{\Delta}}
\def\atd{\atopwithdelims[]}
\def\bm{{\bf m}}


\nref\rGord{B. Gordon, Amer. J. Math. 83 (1961) 363.}
\nref\rAnd{G.E. Andrews, Proc. Nat. Sci. USA 71 (1974) 4082.}
\nref\rAndb{G.E. Andrews, {\it The theory of partitions}
  (Addison-Wesley, London, 1976).}
\nref\rAndq{G.E. Andrews, {\it $q$-series: their development and
 application in analysis, number theory, combinatorics, physics, and
 computer algebra} (American Mathematical Society, Providence, 1986).}
\nref\rRoCa{A.~Rocha-Caridi, in: {\it Vertex operators in mathematics and
 physics},  ed. J.~Lepowsky, S. Mandelstam and I.M. Singer
 (Springer, Berlin, 1985).}
\nref\rModInv{J.L.~Cardy, Nucl.~Phys.~B270 (1986) 186.}
\nref\rBPZ{A.A.~Belavin, A.M.~Polyakov and A.B.~Zamolodchikov,
  Nucl.~Phys.~B241 (1984) 333.}
\nref\rKMM{R.~Kedem, B.M.~McCoy and E.~Melzer,
 Stony Brook preprint, hep-th/9304056.}
\nref\rfctm{E. Melzer,  Int. J. Mod. Phys. A9 (1994) 1115 ~(hep-th/9305114).}
\nref\rWPii{S.O. Warnaar and P.A. Pearce, Melbourne preprint,
 hep-th/9411009.}
\nref\rNone{D. Friedan, Z. Qiu and S.H. Shenker, in:
 {\it Vertex operators in mathematics and physics},
 ed. J.~Lepowsky, S. Mandelstam and I.M. Singer (Springer, Berlin, 1985); \nl
 M.A. Bershadsky, V.G. Knizhnik and M.G. Teitelman, Phys. Lett. B151
 (1985) 31.}
\nref\rKarl{K. Schoutens, Nucl. Phys. B344 (1990) 665.}
\nref\rsupch{B.L. Feigin and D.B. Fuchs, Funct. Anal. Appl. 16 (1982) 114;\nl
 P. Goddard, A. Kent and D. Olive, Commun. Math. Phys. 103 (1986) 105; \nl
 A. Meurman and A. Rocha-Caridi, Commun. Math. Phys. 107 (1986) 263.}
\nref\rAnds{G.E. Andrews, in: {\it The theory and applications of
 special functions}, ed. R. Askey (Academic Press, New York, 1975).}
\nref\rSlat{L.J. Slater, Proc. London Math. Soc. (2) 54 (1952) 147.}
\nref\rBres{D.M. Bressoud, Proc. London Math. Soc. (3) 42 (1981) 478.}
\nref\rKKMMii{R.~Kedem, T.R.~Klassen, B.M.~McCoy and E.~Melzer,
 Phys. Lett. B307 (1993) 68  ~(hep-th/9301046).}
\nref\rtba{Al.B.~Zamolodchikov, Nucl.~Phys.~B342 (1990) 695; \nl
 T.R.~Klassen and E.~Melzer, Nucl.~Phys.~B338 (1990) 485,
 and B350 (1991) 635.}
\nref\rKMflow{T.R.~Klassen and E.~Melzer, Nucl.~Phys.~B370 (1992) 511.}
\nref\rKKMMi{R.~Kedem, T.R.~Klassen, B.M.~McCoy and E.~Melzer,
 Phys. Lett. B304 (1993) 263 ~(hep-th/9211102).}
\nref\rDKMM{S. Dasmahapatra, R. Kedem, B.M. McCoy and E. Melzer,
  J. Stat. Phys. 74 (1994) 239 ~(hep-th/9304150).}
\nref\rFI{P. Fendley and K. Intriligator, Nucl. Phys. 380 (1992) 265.}
\nref\rDynk{F. Ravanini, R. Tateo and A. Valleriani, Int. J. Mod. Phys.
 A8 (1993) 1707 ~(hep-th/9207040).}
\nref\rNew{F. Ravanini, R. Tateo and A. Valleriani, Phys. Lett.
  B293 (1992) 361 ~(hep-th/9207069).}
\nref\rTanig{A.B.~Zamolodchikov, Adv.~Stud.~Pure Math.~19 (1989) 1.}
\nref\rDM{D.A. Depireux and P. Mathieu, Phys. Lett. B308 (1993) 272
 ~(hep-th/9301131).}
\nref\rCIZ{A.~Cappelli, C.~Itzykson and J.-B.~Zuber, Nucl.~Phys.~B280
 (1987) 445.}
\nref\rBaxbook{R.J. Baxter, {\it Exactly solved models in statistical
 mechanics} (Academic Press, London, 1982).}
\nref\rAndp{G.E. Andrews, Scripta Math. 28 (1970) 297.}
\nref\raff{E. Melzer,  Lett. Math. Phys. 31 (1994) 233 ~(hep-th/9312043).}
\nref\rBerk{A. Berkovich, Nucl. Phys. B431 (1994) 315 ~(hep-th/9403073).}
\nref\rQuano{O. Foda and Y.-H. Quano, Melbourne preprints, hep-th/9407191
 and 9408086.}
\nref\rWPi{S.O. Warnaar and P.A. Pearce, Melbourne preprint,
 hep-th/9408136.}
\nref\rBresc{D.M. Bressoud, Quart. J. Math. Oxford (2) 31 (1980) 385.}
\nref\rFNO{B.L. Feigin, T. Nakanishi and H. Ooguri, Int. J. Mod.
 Phys. A7, Suppl. 1A (1992) 217.}
\nref\rBSA{L. Benoit and Y. Saint-Aubin, Int. J. Mod. Phys. A7 (1992) 3023,
 and A9 (1994) 547.}
\nref\rGep{D. Gepner, Caltech preprint, hep-th/9410033.}
\nref\rKRV{J. Kellendonk, M. R\"osgen and R. Varnhagen,
 Int. J. Mod. Phys. A9 (1994) 1009.}
\nref\rIntq{E. Quattrini, F. Ravanini, and R. Tateo,
  Bologna preprint, hep-th/9311116.}

\Title{\vbox{\baselineskip12pt\hbox{TAUP 2211-94}\hbox{hep-th/9412154} }}
{\vbox{\centerline{Supersymmetric Analogs of the} \vskip 13pt
       \centerline{Gordon-Andrews Identities,}  \vskip 13pt
       \centerline{and Related TBA Systems}}}
\centerline{Ezer Melzer~\foot{Work
  supported in part by the US-Israel Binational
  Science Foundation.}  }
\medskip\centerline{\it School of Physics and Astronomy}
\smallskip\centerline{\it Beverly and Raymond Sackler Faculty
  of Exact Sciences}
\smallskip\centerline{\it Tel-Aviv University}
\smallskip\centerline{\it Tel-Aviv 69978, ISRAEL}
\medskip\centerline{email: melzer@ccsg.tau.ac.il}
\vskip 9mm

\centerline{{\bf Abstract}}
\vskip 3mm

The Gordon-Andrews identities, which generalize the
Rogers-Ramanujan-Schur identities, provide product and fermionic
forms for the characters of the minimal conformal field theories
(CFTs) ${\cal M}(2,2k+1)$.  We discuss/conjecture identities of
a similar type, providing two different fermionic forms for the
characters of the models ${\cal SM}(2,4k)$ in the minimal series
of $N$=1 super-CFTs.  These two forms are related to two families
of thermodynamic Bethe Ansatz (TBA) systems, which are argued to be
associated with the $\hat{\phi}_{1,3}^{\rm top}$- and
$\hat{\phi}_{1,5}^{\rm bot}$-perturbations of the models
${\cal SM}(2,4k)$.  Certain other $q$-series identities and
TBA systems are also discussed, as well as a possible
representation-theoretical consequence of our results, based
on Andrews's generalization of the G\"ollnitz-Gordon theorem.

\Date{\hfill}
\vfill\eject

\newsec{Introduction}
\ftno=0

Gordon's theorem~\rGord~in the theory of partitions has an
analytic counterpart due to Andrews~\rAnd\rAndb, which reads as
follows: For $k=2,3,4,\ldots$, ~$s=1,2,\ldots,k$
(and $|q|<1$, which is assumed to hold throughout the paper)
\eqn\GE{ \prod_{n=1 \atop n \not\equiv 0,\pm s({\rm mod}~2k+1)}^\infty
  (1-q^n)^{-1} ~= \sum_{m_1,\ldots,m_{k-1}=0}^\infty
  {q^{N_1^2+N_2^2+\ldots+N_{k-1}^2+N_s+N_{s+1}+\ldots+N_{k-1}} \o
   (q)_{m_1}(q)_{m_2} \ldots (q)_{m_{k-1}}} ~~,}
where
\eqn\Ndef{ N_j~=~m_j+m_{j+1}+\ldots+m_{k-1}~~~~~~~~~~(N_k=0),}
and
\eqn\asub{(z)_0=1~~,~~~~~~~
  (z)_m=(z;q)_{m}=\prod_{\ell=0}^{m-1} (1-z q^\ell)
  ~~~~~~{\rm for}~~m=1,2,3,\ldots.}
To appreciate the broad context in which the Gordon-Andrews identities \GE\
play an important role,
the reader is invited to look at~\rAndq~and references therein.

Using Jacobi's triple product identity (see \eg~eq.~(2.2.10) in~\rAndb),
the lhs of \GE\ can be shown to be equal to $\chi_{1,s}^{(2,2k+1)}(q)$,
where, more generally,
\eqn\GEb{ \chi_{r,s}^{(p,p')}(q) ~=~  \chi_{p-r,p'-s}^{(p,p')}(q)
   ~=~{1\o (q)_\infty} ~\sum_{\ell \in\ZZ}
 \left(q^{\ell(\ell pp'+rp'-sp)}-q^{(\ell p+r)(\ell p'+s)}\right) }
is~\rRoCa~the (normalized) character of the irreducible highest
weight representation of the Virasoro
algebra at central charge ~$c^{(p,p')}=1-{6(p'-p)^2 \o pp'}$~ and
highest weight ~$\Delta_{r,s}^{(p,p')}={(rp'-sp)^2-(p'-p)^2\o 4pp'}$.
For given coprime integers $p$ and $p'$~ ($p'>p\geq 2$) these
representations, with ~$r=1,2,\ldots,p-1$ ~and ~$s=1,2,\ldots,p'-1$,
span the spectrum~\rModInv~of the minimal model
${\cal M}(p,p')$ of conformal field
theory (CFT)~\rBPZ.

Thus the left- and right-hand sides of eq.~\GE\ constitute alternative
expressions,  referred to
as product and fermionic forms, respectively, for the full set of characters
of the minimal CFT ${\cal M}(2,2k+1)$, for all $k\geq 2$.
The form given in \GEb\ is referred to as a bosonic (or free-field)
form. The motivation for the ``physical terminology'' we use is reviewed
in~\rKMM\rfctm. For the most recent developments in the subject of
$q$-series identities, as inspired by work in two-dimensional physics,
see~\rWPii~and references therein.

\medskip
The aim of the present paper is to discuss supersymmetric analogs
of the Gordon-Andrews identities, namely
product-sum identities similar to \GE\ which provide alternative
expressions for the characters of the minimal $N$=1
super-CFTs~\rNone~${\cal SM}(2,4k)$. For various reasons
(see \eg~\rKarl) the family of theories
${\cal SM}(2,4k)$ is the
natural supersymmetric analog of the family ${\cal M}(2,2k+1)$;
this is also revealed by the fermionic forms we will encounter below
for the characters $\hc_{1,s}^{(2,4k)}(q)$ of the super-CFTs.

The (normalized) characters of a generic  $N$=1 superconformal
minimal model $\csm(p,p')$ are given by~\rsupch
\eqn\SGEb{ \hc_{r,s}^{(p,p')}(q) = \hc_{p-r,p'-s}^{(p,p')}(q) =
 {(-q^{\veps_{r-s}})_\infty \o (q)_\infty}
 ~\sum_{\ell\in \ZZ} \left( q^{\ell(\ell pp'+rp'-sp)/2}
          -q^{(\ell p+r)(\ell p'+s)/2}  \right)~,}
where\foot{In order to avoid subtleties related to
 the branches of the square root of $q$,
we restrict ourselves from here on to ~$q\in[0,1)$~ when ~$r-s$~ is even.}
\eqn\epsa{ \veps_a ~=~\cases{
  {1\o 2} &~~~if~ $a$~ is even  ~~($\leftrightarrow$ NS sector) \cr
  1 &~~~if ~$a$~ is odd     ~~~($\leftrightarrow$ ~R~ sector) \cr} }
Here~ $r=1,2,\ldots,p-1$ ~and~ $s=1,2,\ldots,p'-1$ ~as
in the non-supersymmetric case,
but this time $p'>p\geq 2$ are not necessarily coprime
(in fact ${p'-p\o 2}$ and $p$ must be coprime integers).
Depending on the parity of $(r-s)$, the character
\SGEb\ counts the multiplicities of weights in
the irreducible highest-weight representation of
the Neveu-Schwarz (NS) or
Ramond (R) supersymmetric extension of the
Virasoro algebra, with central charge ~$\hat{c}^{(p,p')}=
{3\o 2}(1-{2(p'-p)^2\o pp'})$~ and highest weight
{}~$\hat{\Delta}_{r,s}^{(p,p')}={(rp'-sp)^2-(p'-p)^2\o 8pp'}+
{2\veps_{r-s}-1\o 16}$.

Invoking Jacobi's triple product identity once again,
 the characters of $\csm(2,4k)$ are brought into  the following
product forms:
\eqn\SGEp{ \eqalign{ \hc_{1,s}^{(2,4k)}(q) ~&= ~
  \prod_{n=1 \atop {n\not\equiv 2({\rm mod}~4) \atop
   n\not\equiv 0,\pm s({\rm mod}~4k)}}^\infty (1-q^{n/2})^{-1}
   ~~~~~~~~~~~~~~~~~~~~~~~~~{\rm for}~~s=1,3,\ldots,2k-1,  \cr
  &=~
   \prod_{n=1 \atop n~{\rm odd}}^\infty   (1-q^n)^{-1}
   \prod_{n=1 \atop n \not\equiv 0,\pm s/2({\rm mod}~2k)}^\infty
       (1-q^n)^{-1} ~~~~~{\rm for}~~s=2,4,\ldots,2k-2,  \cr
  &=~
    \prod_{n=1 \atop n \not\equiv 0({\rm mod}~k)}^\infty {1+q^n \o 1-q^n}
   ~~~~~~~~~~~~~~~~~~~~~~~~~~~~~~~~~~~~{\rm for}~~s=2k.  \cr} }
In this connection recall Euler's identity ~$(-q)_\infty=
\prod_{n=1}^\infty (1-q^{2n-1})^{-1}$, which was also used
in the derivation of~\SGEp.

Our main interest is in companion fermionic forms to the products
in eq.~\SGEp. There exist at least
two types of such forms, which will be presented
in the first two subsections of section 2. The $q$-series identities
conjectured there are supplemented in subsection 2.3
by an even stronger conjecture involving two-variable power series.
The two types of fermionic forms are associated with two families of
thermodynamic Bethe Ansatz (TBA) systems. This observation,
as discussed in section 3, helps us identify two
different integrable perturbations of the
models $\csm(2,4k)$ -- one preserving and the other
breaking the $N$=1 supersymmetry -- as corresponding to these TBA systems.
Section 4 contains some further comments and open questions.

\newsec{$q$-series identities}

\subsec{First fermionic form.}

The equality of the fermionic forms of the first type
and the products
on the first line of eq.~\SGEp\ is an analytic counterpart~\rAnds~of
a generalization of the G\"ollnitz-Gordon
theorem (see section 7.4 of~\rAndb). The fermionic sums read,
for ~$s=1,3,\ldots,2k-1$,
\eqn\SGEf{ \eqalign{ \hc_{1,s}^{(2,4k)}&(q) ~=
  \sum_{m_1,\ldots,m_{k-1}=0}^\infty
  {(-q^{1/2})_{N_1} ~q^{{1\o 2}N_1^2+N_2^2+\ldots+N_{k-1}^2
        +N_{(s+1)/2}+N_{(s+3)/2}+\ldots+N_{k-1}} \o
   (q)_{m_1}(q)_{m_2} \ldots (q)_{m_{k-1}}}  \cr
   &=    \sum_{m_1,\ldots,m_{k}=0}^\infty
  {q^{N_1^2+N_2^2+\ldots+N_{k-1}^2
        +N_{(s+1)/2}+N_{(s+3)/2}+\ldots+N_{k-1}-N_1 m_k+{1\o 2}m_k^2} \o
   (q)_{m_1}(q)_{m_2} \ldots (q)_{m_{k-1}}} {N_1 \atd m_k}_q ~~.\cr}}
(In~\rAndb\rAnds~only the case ~$s=2k-1$ ~is presented, with a misprint.)
Here the $N_j$ are as in \Ndef,
and the $q$-binomial coefficient is defined by
\eqn\qbin{ {n \atd m}_q   ~=~
 \cases{ ~{(q)_n \o (q)_m (q)_{n-m}}
  ~~~~~~~~& if ~~$0\leq m \leq n$ \cr  ~0 & otherwise~,\cr} }
for $m,n\in \ZZ$.
The equality between the two lines of \SGEf\ is
established with the aid of identity (3.3.6) of~\rAndb.

Eq.~\SGEf\ provides fermionic forms for all the characters in the
NS sector of $\csm(2,4k)$. In the R sector we are only able to
conjecture similar forms
for two characters, the ones labeled by ~$s=2$~ and ~$s=2k$:
\eqn\Rf{\eqalign{ \hc_{1,2}^{(2,4k)}&(q) ~=
  \sum_{m_1,\ldots,m_{k-1}=0}^\infty
  {(-q)_{N_1}~q^{{1\o 2}N_1(N_1+1)+N_2(N_2+1)+\ldots+N_{k-1}(N_{k-1}+1)} \o
   (q)_{m_1}(q)_{m_2} \ldots (q)_{m_{k-1}}}    \cr
   &=   \sum_{m_1,\ldots,m_{k}=0}^\infty
  {q^{N_1(N_1+1)+N_2(N_2+1)+\ldots+N_{k-1}(N_{k-1}+1)-N_1 m_k
      +{1\o 2}m_k(m_k-1)} \o
   (q)_{m_1}(q)_{m_2} \ldots (q)_{m_{k-1}}} {N_1 \atd m_k}_q ~~,\cr
 \hc_{1,2k}^{(2,4k)}&(q) ~=
   \sum_{m_1,\ldots,m_{k-1}=0}^\infty
   {(-1)_{N_1} ~q^{{1\o 2}N_1(N_1+1)+N_2^2+\ldots+N_{k-1}^2} \o
    (q)_{m_1}(q)_{m_2} \ldots (q)_{m_{k-1}}} \cr
   &=    \sum_{m_1,\ldots,m_{k}=0}^\infty
  {q^{N_1^2+N_2^2+\ldots+N_{k-1}^2
        -N_1 m_k+{1\o 2}m_k(m_k+1)} \o
   (q)_{m_1}(q)_{m_2} \ldots (q)_{m_{k-1}}} {N_1 \atd m_k}_q ~~.\cr}}
These identities have been verified for various small values of $k>2$
up to high orders in the $q$-series, using Mathematica.

The case $k=2$ is already known. Eqs.~\SGEp,\Rf\ reduce then to the
identities
\eqn\SRRSR{ \eqalign{
 \hc_{1,2}^{(2,8)}(q) ~&=
    ~\prod_{n=1 \atop n\not\equiv 0({\rm mod}~4)}^\infty (1-q^n)^{-1}
    ~=~ \sum_{m=0}^\infty {(-q)_m ~q^{m(m+1)/2} \o (q)_m}  \cr
 \hc_{1,4}^{(2,8)}(q) ~&=
    \prod_{n=1 \atop n~{\rm odd}}^\infty {1+q^n \o 1-q^n}
    ~=~ \sum_{m=0}^\infty {(-1)_m ~q^{m(m+1)/2} \o (q)_m}~~,  \cr} }
which are equivalent to identities (8) and (12), respectively, on
Slater's list~\rSlat. For completeness we also write down explicitly
the product-sum identities corresponding to the two
NS characters of the model $\csm(2,8)$, derived from eqs.~\SGEp,\SGEf\
with $k=2$:
\eqn\SRRSNS{ \eqalign{
 \hc_{1,1}^{(2,8)}(q) ~&=
    ~\prod_{n=1 \atop n\not\equiv 0,\pm 1,\pm 2({\rm mod}~8)}^\infty
       (1-q^{n/2})^{-1}
    ~=~ \sum_{m=0}^\infty {(-q^{1/2})_m~ q^{m(m+2)/2} \o (q)_m}  \cr
 \hc_{1,3}^{(2,8)}(q) ~&=
    ~\prod_{n=1 \atop n\not\equiv 0,\pm 2,\pm 3({\rm mod}~8)}^\infty
         (1-q^{n/2})^{-1}
    ~=~ \sum_{m=0}^\infty {(-q^{1/2})_m ~q^{m^2/2} \o (q)_m}~~.  \cr} }
These identities are equivalent to eqs.~(34) and (36) in~\rSlat.
(Eqs.~(38) and (39) of~\rSlat~provide further expressions,
identified as $(-q^{1/2})_\infty \chi_{1,3}^{(3,4)}(-q^{1/2})$ and
$(-q^{1/2})_\infty \chi_{1,1}^{(3,4)}(-q^{1/2})$  for
the same products in \SRRSNS; cf.~also~\rBres.)
Eqs.~\SRRSR--\SRRSNS\ (with $q$ replaced by $q^2$, perhaps) may be called
supersymmetric analogs of the Rogers-Ramanujan-Schur
identities; the latter two celebrated identities are obtained
from \GE\ when specialized to the case $k$=2.

\subsec{Second fermionic form.}

The second type of fermionic forms for the characters of the
same family of models $\csm(2,4k)$ is presented in the
following conjecture:   \nl
For $k=2,3,4,\ldots$ and $s=1,2,\ldots,2k$
\eqn\SGEfc{\eqalign{ \hc_{1,s}^{(2,4k)}(q) ~&=
  \sum_{m_1,\ldots,m_{2k-2}=0}^\infty
  {q^{{1\o 2}(M_1^2+M_2^2+\ldots+M_{2k-2}^2)
        +M_s+M_{s+2}+\ldots+M_{2k-3}} \o
   (q)_{m_1}(q)_{m_2} \ldots (q)_{m_{2k-2}}}
      ~~~~~~~~~~~(s ~{\rm odd}) \cr
  ~&= \sum_{m_1,\ldots,m_{2k-2}=0}^\infty
  {q^{{1\o 2}(M_1^2+M_2^2+\ldots+M_{2k-2}^2)
        +M_s+M_{s+2}+\ldots+M_{2k-2} +{1\o 2}\tilde{M}} \o
   (q)_{m_1}(q)_{m_2} \ldots (q)_{m_{2k-2}}}
      ~~~~~(s ~{\rm even}), \cr} }
where
\eqn\Mdef{ M_j ~=~ m_j +m_{j+1}+\ldots+m_{2k-2}~~,~~~~~~~
   \tilde{M}=m_1+m_3+\ldots+m_{2k-3}~.}
Again, this conjecture has been verified for various small $k>2$ and
many orders in the expansion in powers of $q^{1/2}$. Evidence regarding
the asymptotic behavior of high powers, namely the behavior of the
$q$-series as $q\to 1^-$, will be presented in subsection 3.2.

For $k$=2 the fermionic form \SGEfc\ can be brought by a simple
change of the two summation variables into the fermionic form
on the second lines of eqs.~\SGEf\ and \Rf.
(As noted above, the latter double-sums can be shown to
be equal to the single-sums presented explicitly in
eqs.~\SRRSR--\SRRSNS.) Thus the
identities \SGEfc\ are proven in this case.

It is also worth noting that the special case $s$=3 (still $k$=2)
has already been encountered before~\rKKMMii~in disguise.
To demonstrate that, we first of all note that the $N$=1 super-CFT
$\csm(2,8)$ is in fact equivalent to the model ${\cal M}(3,8)$ in
the $N$=0 series. At the level of characters this equivalence
is exhibited by the decomposition
\eqn\chdec{ \eqalign{
 \hc_{1,1}^{(2,8)}(q)~&=~\chi_{1,1}^{(3,8)}(q)+q^{3/2}\chi_{1,7}^{(3,8)}(q)~~~,
 ~~~~~  \hc_{1,3}^{(2,8)}(q)~=~
     \chi_{1,3}^{(3,8)}(q)+q^{1/2}\chi_{1,5}^{(3,8)}(q)~~,\cr
  \hc_{1,2}^{(2,8)}(q)~&=~\chi_{1,4}^{(3,8)}(q)~~~,~~~~~\hc_{1,4}^{(2,8)}(q)
  ~=~ \chi_{1,2}^{(3,8)}(q)+q\chi_{1,6}^{(3,8)}(q)
  ~=~ 2\chi_{1,2}^{(3,8)}(q)-1~~.\cr} }
Now in~\rKKMMii~fermionic expressions for the characters
$\chi_{1,k+1}^{(3,3k+2)}$ (misprinted there as $\chi_{1,k}^{(3,3k+2)}$)
were conjectured. In the case $k$=2 this conjecture reduces to the
sum representation of $\hc_{1,3}^{(3,8)}$ given in \SGEfc, with the
summation restricted to run over $m_1\in 2\ZZ$. Fermionic forms for the
remaining $\chi_{r,s}^{(3,8)}$ as the sums \SGEfc\ at $k$=2 with
even/odd restrictions on $m_1$ can now be deduced as well using \chdec.

\subsec{A stronger conjecture.}

In section 7.2 of~\rAndb~the generating functions
$J_{k,i}(a;x;q)$ are introduced.
For ~$a=-q^{-1/2}$, in particular, the definition there
leads to
\eqn\Jdef{ \eqalign{ J_{k,i}&(-q^{-1/2};x;q)~
  = ~\sum_{n=0}^\infty  (-x^k)^n q^{kn(n+1)-(i-{1\o 2})n} ~\cr
 &\times ~  {(-q^{1/2})_n~\left( 1+xq^{n+{1\o 2}}-
           (1+q^{n+{1\o 2}})x^i q^{(i-{1\o 2})(2n+1)} \right)
    (-xq^{n+{3\o 2}})_\infty  \o (q)_n ~(xq^{n+1})_\infty}~.\cr}}
We now conjecture that the functions \Jdef,
with $i=1,2,\ldots,k$, admit the following fermionic representations:
\eqn\Jconj{ \eqalign{ J_{k,i}&(-q^{-1/2};x;q)~\cr
   =&      \sum_{m_1,\ldots,m_{k-1}=0}^\infty
  {x^{N_1+N_2+\ldots+N_{k-1}}
   (-q^{1/2})_{N_1} ~q^{{1\o 2}N_1^2+N_2^2+\ldots+N_{k-1}^2
        +N_{i}+N_{i+1}+\ldots+N_{k-1}} \o
   (q)_{m_1}(q)_{m_2} \ldots (q)_{m_{k-1}}}  \cr
  =&  \sum_{m_1,\ldots,m_{2k-2}=0}^\infty
  {x^{M_1+M_3+\ldots+M_{2k-3}}~
   q^{{1\o 2}(M_1^2+M_2^2+\ldots+M_{2k-2}^2)
        +M_{2i-1}+M_{2i+1}+\ldots+M_{2k-3}} \o
   (q)_{m_1}(q)_{m_2} \ldots (q)_{m_{2k-2}}}~~,\cr} }
where the $N_j$ and $M_j$ are defined in \Ndef\ and \Mdef.

At $x=1$ this conjecture reduces to the first equalities in \SGEf\ and
\SGEfc, since it can easily be checked that for $s=1,3,\ldots,2k-1$ eq.~\Jdef\
yields ~$J_{k,(s+1)/2}(-q^{-1/2};1;q)=\hc_{1,s}^{(2,4k)}(q)$,
where the character is given by \SGEb.
Finally, let us note that the same generating functions $J_{k,i}(a;x;q)$
play a similar role also in the non-supersymmetric case. Namely
(see eq.~(7.3.8) in~\rAndb), for $i=1,2,\ldots,k$
\eqn\Jki{ \eqalign{ J_{k,i}&(0;x;q) ~=~
   ~\sum_{n=0}^\infty (-x^k)^n~ q^{(k+{1\o 2})n(n+1)-in}
   ~ {1-x^i q^{i(2n+1)} \o (q)_n ~(xq^{n+1})_\infty}  \cr
   &=      \sum_{m_1,\ldots,m_{k-1}=0}^\infty
  {x^{N_1+N_2+\ldots+N_{k-1}}~
   q^{N_1^2+N_2^2+\ldots+N_{k-1}^2
        +N_{i}+N_{i+1}+\ldots+N_{k-1}} \o
   (q)_{m_1}(q)_{m_2} \ldots (q)_{m_{k-1}}}~~,  \cr}}
so that, in particular, $J_{k,i}(0;1;q) ~=~ \chi_{1,i}^{(2,2k+1)}(q)$.

\newsec{Perturbed CFT, TBA systems, and dilogarithm sum-rules}

The way we arrived at the results described in section 2 was by
considering certain thermodynamic Bethe Ansatz (TBA) systems, which
specify the finite-volume ground state energy of some integrable
massive quantum field theories in 1+1 dimensions as a solution of
a set of nonlinear integral equations~\rtba.
The volume dependence of the ground state energy provides information
about the renormalization group flow of the massive theory from the
CFT in the ultraviolet limit to the massive infrared fixed point.
In particular, the ultraviolet effective central charge ~$\tilde{c}=
c-24\Delta_{\rm min}$, where $c$ is the Virasoro central charge
and $\Delta_{\rm min}$ is the lowest conformal dimension of the CFT,
can be extracted from the TBA system and turns out to be given by
a sum of Rogers dilogarithms. The conformal dimension $\Delta_p$
of the perturbing field, which ``drives'' the theory away from the ultraviolet
fixed point along the renormalization group trajectory, can also be
deduced from that system.

To make the above brief discussion slightly more concrete, let us first
summarize some properties of a generic TBA system of the type of interest
here (see \eg~\rKMflow).
The scaled finite-volume ground state energy is given, as a function of the
scaled dimensionless volume $\rho$,  by
\eqn\erho{ e_0(\rho) ~=~ -{\rho\o 2\pi}~ \sum_{a=1}^n
   \int_{-\infty}^\infty {d\th\o 2\pi}~ \mu_a \cosh\th
     ~\ln\left(1+e^{-\eps_a(\th)}\right)~~,}
where the $\eps_a(\th)$ satisfy the equations
\eqn\tba{ \eps_a(\th) ~=~\rho \mu_a \cosh\th - \sum_{b=1}^n \left(K_{ab}
  \ast \ln(1+e^{-\eps_b}) \right)(\th)~~.}
Here the $\mu_a$ are nonnegative mass parameters, the kernel $K_{ab}(\th)$
is a symmetric matrix of even functions
which depend on the $S$-matrix of the theory, and the convolution is
defined by ~$(f \ast g)(\th) = \int_{-\infty}^\infty {d\th' \o 2\pi}
f(\th-\th')g(\th')$.  The effective central charge obtained from
the above TBA system is
\eqn\ceff{ \tilde{c} ~=~ -12e_0(0) ~=~ {6\o \pi^2} \sum_{a=1}^n
  [{\cal L}(1-x_a) -{\cal L}(1-y_a)]~~,}
where (for $0\leq z\leq 1$)
\eqn\dilog{ {\cal L}(z) ~=~ -{1\o 2} \int_0^z dt \Bigl[ {\ln t\o 1-t}
  +{\ln(1-t) \o t} \Bigr] ~}
is the Rogers dilogarithm and the $x_a, y_a$ are solutions to
the equations
\eqn\xyeq{ 1-x_a ~=~\prod_{b=1}^n x_b^{B_{ab}}~~,~~~~~~~~
   1-y_a ~=~\sigma_a \prod_{b=1}^n y_b^{B_{ab}}~~.}
In \xyeq, $\sigma_a=1$~ if~ $\mu_a=0$~ and 0 otherwise, and
\eqn\Bdef{ B_{ab} ~=~ \delta_{ab}-\int_{-\infty}^\infty {d\th\o 2\pi}
  ~K_{ab}(\th)~~,}
which in all known cases turn out to be rational numbers.

The relation between all that and fermionic
forms of CFT characters is discussed in detail in~\rKKMMii.
Here we just repeat the upshot
of that discussion. Namely, making use of part of the data
in the previous paragraph, construct the fermionic sums
\eqn\fsum{ F_B(q) ~=~ \sum_{m_1,\ldots,m_n=0}^\infty
  q^{{1\o 2} \bm B \bm^t -{\bf A}\cdot \bm}
 \prod_{a=1}^n { (\bm (1-B))_a + u_a \atd m_a}_q ~~,}
where $B$ is the matrix \Bdef, $\bm =(m_1,\ldots,m_n)$,
${\bf A}$ is a real $n$-dimensional vector, and the $u_a$ are some more
parameters; in particular ~$u_a=\infty$~ if $\mu_a$ in \erho--\tba\ is
strictly positive, and otherwise the $u_a$ are such that the upper entries
of all the $q$-binomials in \fsum\ are integral for infinitely many $\bm$
 (note that ${\infty \atd m}_q ={1\o (q)_m}$, and we also
use the convention that ${n\atd m}_q=0$ ~if ~$n\not\in \ZZ$).
Then, letting ~$q=e^{2\pi i\tau}$~ and ~$\tilde{q}=e^{-2\pi i/\tau}$
(with Im$\tau >0$), one has
\eqn\fas{ F_B(q) ~\sim ~ \tilde{q}^{-\tilde{c}/24} ~~~~~~~~~~{\rm as}~~~~~~
  q\to 1^-~~,}
with $\tilde{c}$ given by \ceff--\Bdef. This is the asymptotic behavior
of the characters of a CFT whose effective central charge is $\tilde{c}$,
as dictated by their modular properties.

The papers~\rKKMMii\rKKMMi\rDKMM~suggest that associated with every
integrable (relevant) perturbation of any rational CFT there is a set
of fermionic sums of the type \fsum, containing at least one sum for
each character of that CFT. This set is characterized by a fixed quadratic
form $B$ and by the list of the components of ${\bf u}$ which are infinite.
The remaining components $u_a<\infty$ and the vector ${\bf A}$ depend
on the specific character in question. In particular, in all cases
studied it was observed that the vacuum character (corresponding to the
representation of
lowest highest-weight $\Delta_{\rm min}$ in the CFT) is represented
by \fsum\ with ${\bf A}=0$ and all ~$u_a=0$, whenever ~$u_a<\infty$
(and perhaps some restrictions on the vectors $\bm$ which are summed over).

\subsec{First family of TBA systems.}

The TBA systems which are relevant for the discussion in section
2.1 are obtained by ``folding in half'' the systems derived
in~\rFI~for the most relevant SUSY-preserving perturbations of the even
members in the series of minimal $N$=2 super-CFTs, described by the
Landau-Ginzburg superpotential ~${X^{2k}\o 2k}-\lambda X$
{}~($k=2,3,4,\ldots$). This folding is the supersymmetric analog
of the procedure through which the TBA systems of\foot{Henceforth
we use the symbol ~`${\cal M}+\phi_p$'~ as a shorthand for the phrase
`the $\phi_p$-perturbation of the CFT model ${\cal M}$'.}
{}~${\cal M}(2,2k+1)+\phi_{1,3}$~ are related to the
systems of the theories of $\ZZ_{2k-1}$ parafermions perturbed by the
most relevant thermal operator~\rtba.

Specifically, in both cases the $\ZZ_2 \times \ZZ_{2k-1}$ symmetry
of the unfolded model is reflected in the symmetries
{}~$\mu_a=\mu_{n+1-a}$ ~and ~$K_{ab}(\th)=K_{n+1-a,n+1-b}(\th)$~ of
the data in \tba, where  ($n=2k-2$)~ $n=2k$ ~in the
(non-)supersymmetric case. As a result, the solution of \tba\ satisfies
{}~$\eps_a(\th)=\eps_{n+1-a}(\th)$, and so one finds that
{}~$e_0(\rho)=2e_0^{(2)}(\rho)$, where $e_0^{(2)}(\rho)$ is the scaled
ground state energy of the folded TBA system
\eqn\tbafold{ \eqalign{ e_0^{(2)}(\rho)  ~&=~ -{\rho\o 2\pi}~\sum_{a=1}^{n/2}
   \int_{-\infty}^\infty {d\th\o 2\pi}~ \mu_a \cosh\th
    ~ \ln\left(1+e^{-\eps_a(\th)}\right)~~\cr
   \eps_a(\th) ~&=~\rho \mu_a \cosh\th - \sum_{b=1}^{n/2} \left(K_{ab}^{(2)}
     \ast \ln(1+e^{-\eps_b}) \right)(\th)~~.\cr}}
Here
\eqn\Kfold{ K_{ab}^{(2)}(\th) ~=~ K_{ab}(\th)+K_{a,n+1-b}(\th)~~
 ~~~~~~~~~~~~(a,b=1,2,\ldots,{n\o 2}).}

The analogy with the non-supersymmetric case suggests
that the ~$n=k$~ TBA system, obtained by folding the $N$=2 supersymmetric
system of~\rFI~with $n=2k$, gives the finite-volume ground state energy
of the $N$=1 super-CFT $\csm(2,4k)$ perturbed by the top component
$\hat{\phi}_{1,3}^{\rm top}$ of
the superfield $\hat{\phi}_{1,3}$ (so that the corresponding conformal
dimension is ~$\Delta_p^{(2)}=\hat{\Delta}_{1,3}^{(2,4k)}+{1\o 2}={1\o 2k}$).
This folded TBA system can in principle be -- but has not yet been --
derived from the (nondiagonal) factorizable scattering theory proposed
in~\rKarl~to describe this $N$=1 SUSY-preserving perturbation.

If correct, our proposal implies that the scaled ground
state energy of ~$\csm(2,4k)+\hat{\phi}_{1,3}^{\rm top}$~
is precisely half that of the corresponding
perturbed $N$=2 model, for all ~$\rho\geq 0$.
Two simple consistency
tests support the validity of this relation. First, the effective
central charge
{}~$\tilde{c}=\hat{c}^{(2,4k)}-24\hat{\Delta}_{1,2k-1}^{(2,4k)}={3(k-1)\o 2k}$
{}~of $\csm(2,4k)$ is indeed half the one
{}~$\tilde{c}=c={3\cdot (2k-2)\o (2k-2)+2}={3(k-1)\o k}$~
of the $N$=2 super-CFT.
And second, note that
in the $N$=2 super-CFT the three-point function
$\langle \phi_{\rm min} \phi_p \phi_{\rm min} \rangle$
vanishes, being equal to ~$\langle \phi_p \rangle =0$, while
in the $N$=1 theory it does not, as
$\langle \phi_{\rm min} \phi_p \phi_{\rm min} \rangle \propto
\langle \hp_{1,2k-1}^{(2,4k)} \hp_{1,3}^{(2,4k)} \hp_{1,2k-1}^{(2,4k)}
 \rangle \neq 0$. Consequently~\rtba,
the relation ~$2(1-\Delta_p)=1-\Delta_p^{(2)}$~
between the  dimensions of the corresponding perturbing fields
must hold; and indeed it does, since
{}~$\Delta_p={1\o 2}+{1\o 2\cdot 2k}={2k+1\o 4k}$~ in the $N$=2 theory,
while ~$\Delta_p^{(2)}={1\o 2k}$.

Finally, we point out that the fermionic forms on the second lines
of eqs.~\SGEf\ and \Rf\ follow the construction \fsum\
described above. To see that, note that these fermionic sums
are of the form \fsum, with $n=k$, ~$u_a=\infty$ ~for~$a=1,2\ldots,k-1$,
and $B$ given by
\eqn\Bfold{\eqalign{
  B_{ab}&=(2C_{T_{k-1}}^{-1})_{ab} =2\max(a,b)
    ~~~~~~~~~~~~~~~~~(a,b=1,2,\ldots,k-1),\cr
  B_{ak}&=B_{ka}=2\del_{ak}-1
   ~~~~~~~~~~~~~~~~~~~~~~~~~~~~~~~~~(a=1,2,\ldots,k),\cr}}
where $C_{T_n}$ is the Cartan matrix of the tadpole diagram $A_{2n}/\ZZ_2$
(see~\eg~\rKKMMii). This is precisely what one obtains when plugging
\Kfold\ into \Bdef, where the $K_{ab}(\th)$ are specified in~\rFI~(the
labels ~$a=0,\bar{0}$~ in the latter reference correspond to ~$a=k,k+1$~ in
our notation). The resulting dilogarithm sum-rule for the effective
central charge of the CFT $\csm(2,4k)$, which follows from~\rFI, reads
\eqn\difold{ {6\o \pi^2}~ \left[~ \sum_{a=1}^{k-1} {\cal L}
  \left( {\sin^2 {\pi\o 2k} \o \sin^2{(2a+1)\pi\o 4k} } \right)
  + {\cal L} \left( 1-{1 \o 4\cos^2{\pi\o 4k} } \right)
  - {\cal L}\Bigr({1\o 2}\Bigr) ~\right] ~=~ {3(k-1)\o 2k}~~.}

\subsec{Second family of TBA systems.}

We now briefly discuss a second family of TBA systems which is related
to the second type of fermionic forms for the characters of
the super-CFTs $\csm(2,4k)$, presented in subsection 2.2.
The systems in question
are denoted by~ $T_{2k-2} \diamond T_1$ ~in the notation
of~\rDynk\rNew; the reader is referred
to these references for further details. The effective central charge
extracted from this system is ~$\tilde{c}={3(k-1)\o 2k}$~
(using eq.~(40) of~\rNew~with ~$r=2k-2$~ and ~$h=4k-3$, as
appropriate for $T_{2k-2}$). Furthermore, using the so-called
periodicity ~$P={4k\o 4k-3}$~ of the system we read off
the conformal dimension of the perturbing field ~$\Delta_p =
1-{2\o P}= -{2k-3\o 2k}$, which applies when
the three-point function
$\langle \phi_{\rm min} \phi_p \phi_{\rm min} \rangle$
in the ultraviolet CFT is nonvanishing.

Noting that the effective central charge is that of $\csm(2,4k)$,
and that $\Delta_p=\hat{\Delta}_{1,5}^{(2,4k)}$
is the dimension of the bottom component $\hp_{1,5}^{\rm bot}$
of the superfield $\hat{\phi}_{1,5}$ in this CFT
(and that indeed $\langle \phi_{\rm min} \phi_p \phi_{\rm min} \rangle \propto
\langle \hp_{1,2k-1}^{(2,4k)} \hp_{1,5}^{(2,4k)} \hp_{1,2k-1}^{(2,4k)} \rangle
\neq 0$), we propose that the TBA system
{}~$T_{2k-2} \diamond T_1$ ~is associated with
{}~$\csm(2,4k)+\hat{\phi}_{1,5}^{\rm bot}$.
An indication for the integrability of
this SUSY-breaking perturbation is revealed
by applying Zamolodchikov's counting argument~\rTanig, which shows
the existence of a nontrivial integral of motion of spin $s$=3 for all
$k$ (for $k$=2 also $s$=5,7 are found).\foot{The $\hp_{1,5}^{\rm
bot}$-perturbation is not included in the set of integrable
perturbations of $\csm(p,p')$ discussed in~\rDM, perhaps
because it is irrelevant in the case of the unitary series
$\csm(p,p+2)$, namely ~$\hD_{1,5}^{(p,p+2)}>1$.}

The proposal above is apparently new for all $k>2$. In the case
$k$=2 it is consistent with the identification noted in~\rNew~of
the system ~$T_{2} \diamond T_1$ ~as corresponding to
{}~${\cal M}(3,8)+\phi_{1,3}$; consistency
is ensured here by the equivalence
(see subsection 2.2) of the latter theory
with ~$\csm(2,8)+\hp_{1,5}^{\rm bot} =
\csm(2,8)+\hp_{1,3}^{\rm bot}$.

For completeness, let us further suggest that the TBA system
{}~$T_{2k-1} \diamond T_1$ ~with $k\geq 2$ is associated with
the perturbation of the tensor-product CFT
{}~${\cal M}(3,4) \otimes {\cal M}(2,2k+1)$~ by the operator
{}~$\phi_{1,3}^{(3,4)} \otimes \phi_{1,2}^{(2,2k+1)}$.\foot{The first
model in this series, with $k$=2, is equivalent to
{}~${\cal M}(5,12)+\phi_{2,7}$ ~(with the $E_6$
modular-invariant partition function~\rCIZ~for ${\cal M}(5,12)$).}
This identification is supported by the match between the
effective central charge and the dimension $\Delta_p$ of the perturbation
in this theory with the ones deduced from the TBA system.
(Note that in this case
$\langle \phi_{\rm min} \phi_p \phi_{\rm min} \rangle =
\langle \phi_{1,3}^{(3,4)} \rangle \langle \phi_{1,k}^{(2,2k+1)}
\phi_{1,2}^{(2,2k+1)} \phi_{1,k}^{(2,2k+1)} \rangle =0$,
and therefore $\Delta_p$ is related to the periodicity $P$ of the system
according to the choice ~$\Delta_p=1-{1\o P}$~ in eq.~(3.18) of~\rDynk.)

The TBA systems
{}~$T_{n} \diamond T_1$ ~are completely massive, in the sense
that all $\mu_a>0$ in \tba, and hence
the corresponding factorizable scattering theories have
diagonal $S$-matrices. Their scattering amplitudes are given in~\rNew.
Thus the proposals above also answer the question which
perturbed CFTs all these diagonal $S$-matrix theories describe.

Finally, it is straightforward to see that the fermionic sums
\SGEfc\ are of the type obtained through the construction \fsum\
based on the system ~$T_{2k-2} \diamond T_1$: it follows
from \Bdef\ and~\rDynk\rNew~that
the matrix $B$ in this case is ~$B=C^{-1}_{T_{2k-2}} \otimes C_{T_1}
=C^{-1}_{T_{2k-2}}$~ (\ie~ $B_{ab}={\rm max}(a,b)$, and so
{}~${1\o 2}\bm B \bm^t ={1\o 2}(M_1^2 +M_2^2+\ldots +M_{2k-2}^2)$),
and ~$u_a$=$\infty$~ for all ~$a=1,2,\ldots,2k-2$.
The corresponding dilogarithm sum-rule for $\tilde{c}$
is a special case of the general results in~\rDynk\rNew. The implied
asymptotic behavior \fas\ of the conjectured fermionic forms in
\SGEfc\ provides further support for their correctness.

\newsec{Comments and open questions}

\subsec{Regarding proofs.}

There are several ways to approach the problem
of proving the conjectures in section 2
(see~\rAndb\rAndq~and references
therein for some of the known methods).
An approach which has a physical interpretation
is based on a remarkable connection between $q$-series representing
CFT characters and one-dimensional configuration sums
encountered in corner transfer matrix computations
in exactly solvable lattice models~\rBaxbook. In the latter
framework the infinite $q$-series arise as limiting cases of
families of polynomials,  termed finitized
characters in~\rfctm. It is sometimes easier to prove the
stronger result of equality between
finitized versions of bosonic and fermionic sum forms of the
characters, from which the $q$-series identities
follows~\rfctm [29-33].

I believe one can find and prove polynomial identities
which imply the $q$-series identities conjectured in section 2.
It would be interesting to see whether the corresponding
polynomials are in fact one-dimensional configuration sums
of any solvable lattice models.

\subsec{Combinatorial interpretation.}

As mentioned in the introduction, $q$-series identities of the
type discussed in this paper can be viewed as analytic counterparts of
combinatorial theorems in the theory of partitions~\rAndb.
In particular, according to theorem 7.11 in~\rAndb~and the first
line of eq.~\SGEp, the Neveu-Schwarz characters $\hc_{1,2i-1}^{(2,4k)}$
{}~($i=1,2,\ldots,k$) are the generating functions
\eqn\comb{ \hc_{1,2i-1}^{(2,4k)}(q)~=~\sum_{n=0}^\infty
   D_{k,i}(n)~ q^{n/2}~~~~~~~~~~~~~~~~(i=1,2,\ldots,k)}
of the number of partitions $D_{k,i}(n)$ of ~${n\o 2}\in
{1\o 2}\ZZ_{\geq 0}$~ in the form ~${n\o 2}=b_1+b_2+\ldots+b_m$
 ~($b_j \in {1\o 2}\ZZ_{\geq 1}$) in which no half-odd-integral part
is repeated, ~$b_j\geq b_{j+1}$, ~$b_j-b_{j+k-1}\geq 1$~ if
{}~$b_j\in \ZZ+{1\o 2}$, ~$b_j-b_{j+k-1}> 1$~ if
{}~$b_j\in \ZZ$, and at most ~$i-1$~ parts are $\leq 1$.

The first interesting problem which arises is to show
that the fermionic forms in eq.~\SGEf\ and on the first line of
eq.~\SGEfc\ can be directly interpreted as the generating functions
of the coefficients $D_{k,i}(n)$ (analogous result pertaining
to the rhs of eq.~\GE\ is found in~\rBresc). It is furthermore
interesting to explore possible combinatorial interpretations
of the Ramond characters $\hc_{1,2i}^{(2,4k)}$, for which fermionic
forms are conjectured in \Rf\ and the second line of \SGEfc.

Next, by analogy with the results of~\rFNO~(theorem 3.6)
in the non-supersymmetric case,
one may suspect that the combinatorial interpretation described above
for the Neveu-Schwarz characters
has the following representation-theoretical consequence.
Let $v_{k,i}$ be a highest-weight state of conformal weight
$\hD_{1,2i-1}^{(2,4k)}$ in the Verma module of the Neveu-Schwarz
supersymmetrically extended Virasoro algebra at central
charge $\hat{c}^{(2,4k)}$. Then the set of states
\eqn\set{ W_{-b_1} W_{-b_2}\ldots W_{-b_m} v_{k,i}~~}
form a basis for the {\it irreducible} highest-weight representation.
Here ~$W_b=L_b$~ if ~$b\in\ZZ$, ~$W_b=G_b$~ if ~$b\in\ZZ+{1\o 2}$~
(using standard notation for the generators of the $N$=1
super-Virasoro algebra), and the $b_j$ are as in the above definition of the
$D_{k,i}(n)$. To prove this statement, it is sufficient -- by
virtue of \comb\ -- to show that the set \set\ is linearly
independent modulo singular states in the Verma module (in this connection
the results of~\rBSA~are important).

\subsec{Some more identities.}

In subsection 2.1, eqs.~\SGEf\ and \Rf, we noted the equality of
fermionic $k$-multiple  sums of the form \fsum\ (which according
to~\rKKMMii~carry a natural physical interpretation in terms of
fermionic quasiparticles) with ($k-1$)-multiple sums of a slightly
modified form. In particular, in the case of $k$=2 the modified
single-sums, representing the characters \SRRSR--\SRRSNS\
of the $N$=1 super-CFT $\csm(2,8)$, appear on Slater's
list~\rSlat~in the product-sum identities (8), (12), (34),
and (36). It is amusing to go through the list and try to find
and interpret other modified fermionic sums as representing
characters of some super-CFTs.

We found two sets of such sums, described below. Curiously,
together with the model $\csm(2,8)$ the
corresponding two super-CFTs $\csm(3,5)$ and $\csm(3,7)$ exhaust
all the $N$=1 supersymmetric minimal models which are
equivalent to non-supersymmetric ones:
\eqn\modeleq{ \csm(2,8)~\simeq~{\cal M}(3,8)~~,~~~
              \csm(3,5)~\simeq~{\cal M}(4,5)~~,~~~
              \csm(3,7)~\simeq~{\cal M}_{E_6}(7,12)~~}
(the subscript in the last model indicates the type of
modular invariant partition function, according to the classification
in~\rCIZ). The decomposition of characters of the two
super-CFTs, analogous to \chdec\ for $\csm(2,8)$, reads
\eqn\deci{  \eqalign{
 \hc_{1,1}^{(3,5)}~&=~\chi_{1,1}^{(4,5)}+q^{3/2}\chi_{1,4}^{(4,5)}~~~,
 ~~~~~  \hc_{1,3}^{(3,5)}~=~
     \chi_{1,2}^{(4,5)}+q^{1/2}\chi_{1,3}^{(4,5)}~~,\cr
  \hc_{1,2}^{(3,5)}~&=~\chi_{2,2}^{(4,5)}~~~,
  ~~~~~~~~~~~~~~~~~~~~~\hc_{1,4}^{(3,5)}~=~ \chi_{2,1}^{(4,5)}~~,\cr} }
and
\eqn\decii{  \eqalign{
 \hc_{1,1}^{(3,7)}~&=~\chi_{1,1}^{(7,12)}+q^{3/2}\chi_{1,5}^{(7,12)}
           + q^4 \chi_{1,1}^{(7,12)}+q^{25/2}\chi_{1,11}^{(7,12)}~~,\cr
 \hc_{1,3}^{(3,7)}~&=~\chi_{3,5}^{(7,12)}+q^{1/2}\chi_{3,7}^{(7,12)}
           + q^{5/2} \chi_{3,1}^{(7,12)}+q^5 \chi_{3,11}^{(7,12)}~~,\cr
 \hc_{1,5}^{(3,7)}~&=~\chi_{2,5}^{(7,12)}+q^{1/2}\chi_{2,1}^{(7,12)}
           + q^{3/2} \chi_{2,7}^{(7,12)}+q^8 \chi_{2,11}^{(7,12)}~~,\cr
 \hc_{1,2}^{(3,7)}~&=~\chi_{2,4}^{(7,12)}+q^{3} \chi_{2,8}^{(7,12)}~~,\cr
 \hc_{1,4}^{(3,7)}~&=~\chi_{3,4}^{(7,12)}+q \chi_{3,8}^{(7,12)}~~,\cr
 \hc_{1,6}^{(3,7)}~&=~\chi_{1,4}^{(7,12)}+q^{5} \chi_{1,8}^{(7,12)}~~.\cr} }

Now substituting in \deci\
the fermionic forms of the characters $\chi_{r,s}^{(4,5)}$,
which were conjectured in~\rKKMMii~and proven in~\rfctm~(see also~\rBerk\rGep),
simple use of identity (3.3.6) of~\rAndb~leads to
\eqn\TIMf{ \eqalign{
 \hc_{1,1}^{(3,5)}(q) ~&=~ \sum_{m=0 \atop m~{\rm even}}^\infty
  { (-q^{1/2})_{m/2} ~q^{3m^2/8}  \o (q)_m} ~~~~,\cr
 \hc_{1,3}^{(3,5)}(q) ~&=~ \sum_{m=0 \atop m~{\rm even}}^\infty
  { (-q^{1/2})_{m/2} ~q^{m(3m-4)/8}  \o (q)_m}
   ~=~ \sum_{m=1 \atop m~{\rm odd}}^\infty
  { (-q^{1/2})_{(m+1)/2} ~q^{3(m^2-1)/8}  \o (q)_m} ~~,\cr
 \hc_{1,2}^{(3,5)}(q) ~&=~     \sum_{m=1 \atop m~{\rm odd}}^\infty
  { (-q)_{(m-1)/2} ~q^{(m-1)(3m-1)/8}  \o (q)_m}
   ~=~  \sum_{m=0 \atop m~{\rm even}}^\infty
  { (-q)_{m/2} ~q^{m(3m-2)/8}  \o (q)_m}   ~~,\cr
 \hc_{1,4}^{(3,5)}(q) ~&=~     \sum_{m=1 \atop m~{\rm odd}}^\infty
  { (-q)_{(m-1)/2} ~q^{3(m^2-1)/8}  \o (q)_m}~~~.} }
The first sums in each of the four lines of~\TIMf~appear in eqs.~(100),
(95), (62), and (63) of~\rSlat, respectively. To see that, replace $q$ by
$q^2$ in the first two lines of~\TIMf, change the summation variable
$m$ to $2n$ ($2n-1$) when $m$ is even (odd), and correct the misprint
in eq.~(100) of~\rSlat, where the factor $(-q^2; q^2,n)$ should read
$(-q; q^2,n)$. The equalities between the pair of sums on lines
two and three of \TIMf\ is apparently not noted in~\rSlat.
Product forms for all the characters of $\csm(3,5)$ can be inferred
from the corresponding identities in~\rSlat~(see also~\rKRV).

Turning next to the model $\csm(3,7)$, we speculate that the
sums in eqs.~(118), (117), and (119) of~\rSlat, respectively,
provide  the following
fermionic forms for the three Neveu-Schwarz characters
of this model:
\eqn\fsla{ \eqalign{
  \hc_{1,1}^{(3,7)}(q) ~&=
   ~ \sum_{m=0 \atop m~{\rm even}}^\infty
      {(-q^{1/2})_{m/2}~ q^{m(m+4)/8} \o (q)_m}~~~,~~~~~\cr
  \hc_{1,3}^{(3,7)}(q) ~&=
   ~ \sum_{m=0 \atop m~{\rm even}}^\infty
      {(-q^{1/2})_{m/2}~ q^{m^2/8} \o (q)_m}~~~,\cr
  \hc_{1,5}^{(3,7)}(q) ~&=
   ~ \sum_{m=1 \atop m~{\rm odd}}^\infty
      {(-q^{1/2})_{(m+1)/2}~ q^{(m-1)(m+3)/8} \o (q)_m}~~~,\cr}}
while the sums in eqs.~(81), (80), and (82) are equal to the three
Ramond characters:
\eqn\fslb{ \eqalign{
  \hc_{1,2}^{(3,7)}(q) ~&=
   ~ \sum_{m=0 \atop m~{\rm even}}^\infty
      {(-q)_{m/2}~ q^{m(m+2)/8} \o (q)_m}~,~~~\cr
  \hc_{1,4}^{(3,7)}(q) ~&=
   ~ \sum_{m=1 \atop m~{\rm odd}}^\infty
      {(-q)_{(m-1)/2}~ q^{(m^2-1)/8} \o (q)_m}~,\cr
  \hc_{1,6}^{(3,7)}(q) ~&=
   ~ \sum_{m=1 \atop m~{\rm odd}}^\infty
      {(-q)_{(m-1)/2}~ q^{(m-1)(m+5)/8} \o (q)_m}~~~.\cr}}
One way to prove this would be to show the equality of the product
forms implied by the identities in~\rSlat~listed above and the
free-field forms~\SGEb~of the corresponding characters.
Unfortunately, unlike in other cases,
 it seems that a more powerful tool than Jacobi's
triple product identity is required for that purpose.

\subsec{Some more TBA systems.}

Like~\rKKMMii\rKKMMi, section 3 demonstrates the fruitful interplay
between $q$-series identities related to CFT characters and TBA systems.
Here we would like to describe some further
observations regarding TBA systems of
the type encountered in subsection 3.2,
with a few comments on their relation to $q$-series.
Again, we use the notation introduced in~\rDynk.

The family of TBA systems ~$T_n \diamond T_1$~ discussed in subsection
3.2 can be embedded in the general set of systems
{}~$(T_n \diamond T_k)_\ell$.
Here ~$n,k=1,2,\ldots$~ and ~$\ell=1,2,\ldots,k$;
when $k$=1 the redundant index $\ell$=1 is suppressed.
The identification of the perturbed CFTs corresponding to these
systems, as well as to their ``unfolded''
partners ~$(A_{2n} \diamond T_k)_\ell$  (see subsection 3.1),
was not attempted
in~\rDynk\rNew\rIntq.\foot{In~\rNew~attention was restricted to
{}~$(G\diamond T_k)_\ell$~ with $h_G$ even, whereas in our case
{}~$h_{A_{2n}}=h_{T_n}=2n+1$~ is odd.}
Using the dilogarithm sum-rules in~\rIntq, the following formula for the
effective central charges of the ultraviolet CFTs is obtained:
\eqn\cmany{ \eqalign{
  \tilde{c}((T_n \diamond T_k)_\ell) ~&=~
   {\textstyle {1\o 2}} \tilde{c}((A_{2n} \diamond T_k)_\ell) \cr
  &=~n\left( {k(2k+1)\o 2(n+k+1)} -{\ell(\ell-1)\o 2n+\ell+1}
   -{(k-\ell)(2k-2\ell+1) \o 2(n+k-\ell+1)} \right)~~.\cr}}
In addition, the conformal dimension of the perturbing field, as obtained
from the periodicity ~$P={2(n+k+1)\o 2n+1}$~ of the system, is
\eqn\Dmany{ \Delta_p ~=~ \cases{ 1-{1\o P}~=~{2k+1\o 2(n+k+1)}
    &~~~~if~~~ $\langle \phi_{\rm min} \phi_p \phi_{\rm min} \rangle =0$,\cr
 1-{2\o P}~=~{k-n\o n+k+1} &~~~~if~~~
  $\langle \phi_{\rm min} \phi_p \phi_{\rm min} \rangle \neq 0$.\cr} }

A complete analysis of the general $(n,k,\ell)$ case seems formidable.
In the sequel we extend the system identifications proposed in
subsection 3.2 in a few special cases:

\no $\bullet$ ~$A_{2n} \diamond T_1$:
{}~Here we suspect that the
unperturbed CFT is unitary, implying that ~$c=\tilde{c}={3n\o n+2}$~
and that the first choice in eq.~\Dmany\ applies, yielding
{}~$\Delta_p={3\o 2(n+2)}$. The corresponding CFT is presumably a
special point along the continuous line of theories
obtained from the $SU(2)_n$ WZW model by a marginal perturbation.
In particular, for $n$=1 we find $c$=1 and ~$\Delta_p={1\o 2}$,
and the corresponding finite-volume ground state energy is that
of two free massive particles of equal masses;
the unperturbed CFT is the so-called free Dirac point along
the $c$=1 gaussian line. This is consistent
with the fact that the folded system ~$T_1 \diamond T_1$~
is associated~\rNew~with the thermal perturbation of the Ising model.

\no $\bullet$ ~$(T_1 \diamond T_k)_1$: ~In this case
{}~$\tilde{c}=1-{6\o 2(k+1)(k+2)}$, identified as the effective central
charge of the minimal model ${\cal M}(k+2,2k+2)$ when $k$ is odd,
and ${\cal M}(k+1,2k+4)$ when $k$ is even. Using the first (second) choice
in~\Dmany\ for odd (even) $k$, the perturbing field is identified
as $\phi_{2,1}^{(k+2,2k+2)}$ ($\phi_{1,5}^{(k+1,2k+4)}$).
It can be
checked that these choices are consistent with the condition on the
three-point function stated in \Dmany,
using the fact that
{}~$\phi_{\rm min}^{(k+2,2k+2)}=\phi_{(k+1)/2,k}^{(k+2,2k+2)}$~
{}~($\phi_{\rm min}^{(k+1,2k+4)}=\phi_{k/2,k+1}^{(k+1,2k+4)}$)~
for $k$ odd (even) and the known fusion rules of the minimal models.

Of special interest is the case $k$=2, corresponding to
{}~${\cal M}(3,8)+\phi_{1,5}$. Omitting the
technical details, we state that the system
{}~$(T_1 \diamond T_2)_1$~ can be shown to be equivalent to
the $k$=2 folded TBA system  in subsection 3.1.
(By the same token, the unfolded system
derived in~\rFI~for the most relevant
perturbation of the second model in the  $N$=2 minimal series
is just  ~$(A_2 \diamond T_2)_1$.) A hint to this fact is seen
by noting that for $k$=2 the matrix $B$ of \Bfold\ is simply
{}~$C_{T_2}=(C_{T_1})^{-1}\otimes C_{T_2}$. Now the system in
subsection 3.1 was associated  with
{}~$\csm(2,8)+\hp_{1,3}^{\rm top}$.
As seen from \chdec, this theory
is indeed equivalent to ~${\cal M}(3,8)+\phi_{1,5}$,
in line with our identification here.

As another special model in the series we note that
{}~$(T_1\diamond T_5)_1$~ corresponds to ~${\cal M}(7,12)+\phi_{2,1}$.
Using the equivalence in \modeleq\ this perturbed theory is
related to ~$\csm(3,7)+\hp_{1,5}^{\rm top}$. The TBA system for
another SUSY-preserving
perturbation of the same model $\csm(3,7)$, namely by
$\hp_{1,3}^{\rm top}$, is identified in~\rDynk~as
{}~$(A_1\diamond T_2)_2$. When used in the construction \fsum, the
latter system gives rise to the fermionic forms \fsla--\fslb\
for the characters of $\csm(3,7)$ (once the factors $(-q^\eps)_n$
in \fsla--\fslb\ are expanded using eq.~(3.3.6) of~\rAndb).

\no $\bullet$ ~$(T_1 \diamond T_k)_2$~ with odd ~$k\geq 3$:
{}~Here \cmany\ gives ~$\tilde{c}={3\o 5}+(1-{6\o k(k+2)})$,
leading to the identification of the unperturbed CFT as
{}~${\cal M}(3,5)\otimes {\cal M}(k,k+2)$. Furthermore,
$\Delta_p={2k+1\o 2(k+2)}=\Delta_{2,1}^{(3,5)}+
\Delta_{1,2}^{(k,k+2)}$~ using the first choice in \Dmany,
which is indeed the appropriate one for the perturbing field
$\phi_{2,1}^{(3,5)}\otimes \phi_{1,2}^{(k,k+2)}$.

\bigskip
To provide stronger support for the above pairing
of TBA systems and perturbed CFTs, as well as
those in section 3, one
must compare the (numerical) solution of the TBA systems with
results for the finite-volume ground state energies obtained
from conformal perturbation theory, as in~\rtba.
Another direction for future work based on the discussion
in the present subsection, is to look for fermionic forms
for the characters of the encountered unperturbed CFTs,
following the general construction \fsum.

\vfill\eject
\listrefs

\bye\end